\documentclass{article}

\usepackage{cite}
\usepackage{geometry}
\usepackage[T1]{fontenc}
\usepackage[dvipdfmx]{graphicx,color}
\usepackage{authblk}

\hypersetup{
setpagesize=false,
 bookmarksnumbered=true,
 bookmarksopen=true,
 colorlinks=true,
 linkcolor=blue,
 citecolor=red,
}

\title{Effect of optical crosstalk on performance of ILD AHCAL}
\author{Linghui Liu\thanks{Talk presented at the International Workshop on Future Linear Colliders (LCWS2017), Strasbourg, France, 23--27 October 2017. C17-10-23.2. \href{mailto:linghui@icepp.s.u-tokyo.ac.jp}{\texttt{linghui@icepp.s.u-tokyo.ac.jp}}}}

\affil{Department~of~Physics,~Graduate~School~of~Science,~The~University~of~Tokyo 7-3-1,~Hongo,~Bunkyo-ku,~Tokyo,~113-0033~Japan}
\date{}

\begin{document}
\maketitle

\begin{abstract}
The design of the ILD AHCAL is optimized to make the best use of the PFA and its active layers are composed of about eight million scintillator tiles readout by SiPMs.
If there is optical crosstalk between the tiles, the calorimeter performance can be worsened.
Simulation studies on the effect of the optical crosstalk on the calorimeter performance were performed to define the requirement for the active layer design.
\end{abstract}

\section{Introduction}
The International Linear Collider (ILC) is a next-generation linear electron-positron collider planned to be constructed to search for new physics.
The International Large Detector (ILD) \cite{TDR4} is one of the detector concepts for the ILC detector.
In order to achieve high jet energy resolution, the design of the ILD is optimized for the reconstruction method called "Particle Flow Algorithm (PFA)" \cite{PFA}.
The PFA distinguishes each particle in jets and measures the energy with the most appropriate detector according to the particle type.
To distinguish each particle in the detector, hit clusters in calorimeters have to be well separated.
Therefore the PFA requires highly granular calorimetry.
In the baseline design for the analog hadron calorimeter (AHCAL) of the ILD, the active layers are composed of aligned 30~mm~$\times$~30~mm scintillator tiles and silicon photomultipliers (SiPM) to readout the scintillation light.
But if there is some optical crosstalk between the scintillator tiles, the clusters in the shower are smeared by the false hits due to the optical crosstalk, hence the jet energy resolution is worsened.
The effect of the optical crosstalk between tiles on the AHCAL performance was studied by means of MC simulations.

\section{Optical crosstalk in different active layer designs}
In addition to the baseline design, alternative active layer designs such as megatile are also being studied to improve the performance.
Some of the new designs may have a higher optical crosstalk.

In the baseline design, each scintillator tile is wrapped with a reflector foil and readout by a SiPM placed at the center of the tile.
The optical crosstalk between neighboring tiles in the baseline design was measured to be below 1\% in a cosmic ray test.

\begin{figure}[h!]
 \centering
 \includegraphics[width=70mm]{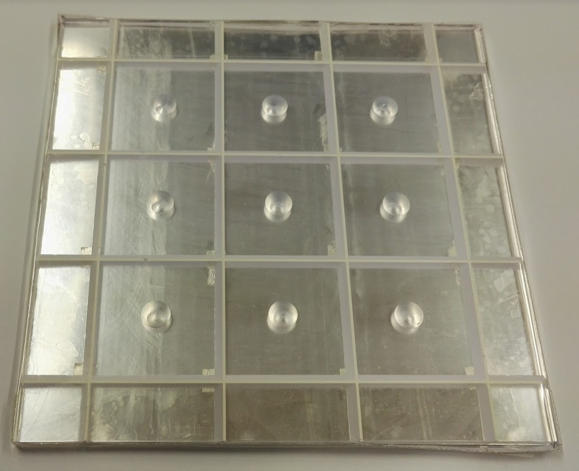}
 \caption{Prototype of megatile \cite{megatile}.}
 \label{figure:megaproto}
\end{figure}

The megatile design is based on a large scintillator plate optically divided into 30~mm$\times$30~mm tiles.
This design can reduce the production cost and simplify the tile assembly.
On the other hand, it may suffer from a higher optical crosstalk.
The optical crosstalk of the megatile design was measured in a prototype shown in Fig. \ref{figure:megaproto} to be 3--5\% \cite{megatile}.

\section{Simulation}
In this study the effect on the jet energy resolution was quantitatively studied by a simulation based on the standard iLCSoft \cite{iLCSoft}.

\subsection{Simulation software}
For the simulation, we used iLCSoft, a common software framework for linear collider studies.
In the iLCSoft, the events are generated in a detector model constructed in a new framework called DD4hep \cite{dd4hep}.
Each hit is digitized in the detectors and the events are reconstructed and analyzed using the PFA.
The version of the iLCSoft used in this study was v01-19-02 and the version of the detector model was ILD\_l1\_v01.

Digitization and reconstruction process have some parameters regarding detector responses.
These parameters are calibrated using $\mu^-, \gamma$ and $K^0_L$ with known energies.
  
\subsection{Simulation setup}
To estimate the jet energy resolution, light quark (uds) pairs were generated and injected into the detectors from the central interaction point.
Only events where particles were injected in the AHCAL main barrel ($|\cos\theta|<0.7$) were used for the jet energy resolution estimation.

\begin{figure}[h!]
 \centering
 \includegraphics[width=120mm]{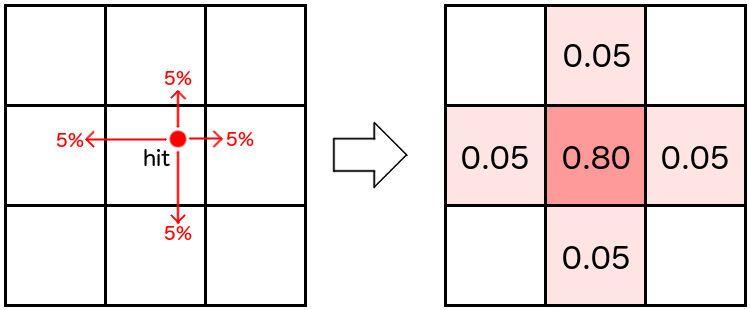}
 \caption{Implementation of the optical crosstalk.}
 \label{figure:xtalkproc}
\end{figure}

We developed a new processor to implement optical crosstalk into the standard iLCSoft.
In this processor which works before digitization, a small fixed fraction of the original hit energy is given to neighboring tiles as shown in Fig. \ref{figure:xtalkproc}.
We assumed a uniform constant crosstalk only to the four neighboring tiles and didn't consider the secondary crosstalk.
The applied crosstalk fractions range from 1\% to 20\%.
After applying the crosstalk we performed normal digitization and reconstruction and investigated the effect of the crosstalk on the jet energy resolution.

\begin{figure}[h!]
 \centering
 \includegraphics[width=120mm]{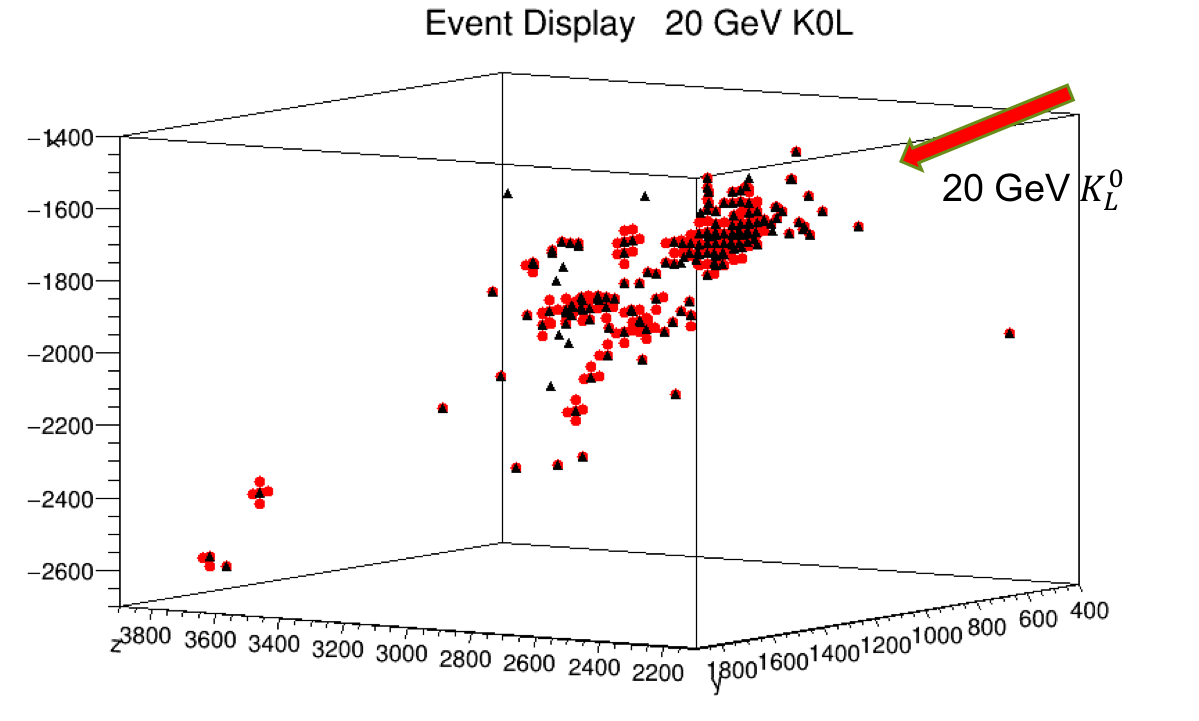}
 \caption{Hit positions after digitization. Black points are hits without crosstalk, red points are hits with 10\% crosstalk.}
 \label{figure:xtalkdisplay}
\end{figure}

Fig. \ref{figure:xtalkdisplay} shows calorimeter hits generated when a 20~GeV $K^0_L$ is injected.
The black points show the hits when no crosstalk is applied while the red points show the hits when 10\% crosstalk is applied.

It can be seen that the cluster shape is broadened and the energy distribution is changed by the crosstalk.
This might affect the performance of the clustering in the PFA and the software compensation discussed in Sec. \ref{section:SC}.
If the energy of the original hit is rather small, the leaked energy and the original hit can vanish because the energy is below hit trigger threshold which is set to be 0.5~MIP in this study.
These decreases of the energy detection efficiency make the energy resolution worse.

\subsection{Software compensation}
\label{section:SC}
Hadronic showers include electromagnetic sub-showers from photons or electrons generated during the shower development.
Generally the detection efficiency of hadrons is worse than that of electrons or photons due to the invisible energy carried by neutrinos, nuclear recoil and excitation.
The energy fraction of the electromagnetic component in a hadronic shower strongly fluctuates from shower to shower, thus this detection efficiency difference has to be compensated for precise energy measurement.

The Pandora PFA makes this correction with an algorithm called "software compensation (SC)" \cite{SC}.
Software compensation exploits the fact that electromagnetic showers tend to have higher energy density than purely hadronic showers.
One can distinguish the shower type using this energy density difference.
Practically software compensation does not strictly determine the shower type, but re-weight each hit according to its energy density.
The weight function is defined as:
\begin{equation}
w(\rho) = p_1\exp(p_2\rho)+p_3 \label{eqn:weight}
\end{equation}
while $p_i$ ($i=1,2,3$) are the parameters dependent on the shower energy:
\begin{eqnarray}
p_1 &=& p_{10}+p_{11}\times E+p_{12}\times E^2\\
p_2 &=& p_{20}+p_{21}\times E+p_{22}\times E^2\\
p_3 &=& \frac{p_{30}}{p_{31}+e^{p_{32}\times E}}.
\end{eqnarray}
The parameters are determined based on the observed energy dependence in experiments.

\begin{figure}[h!]
 \begin{minipage}{0.48\hsize}
  \centering
  \includegraphics[height=60mm]{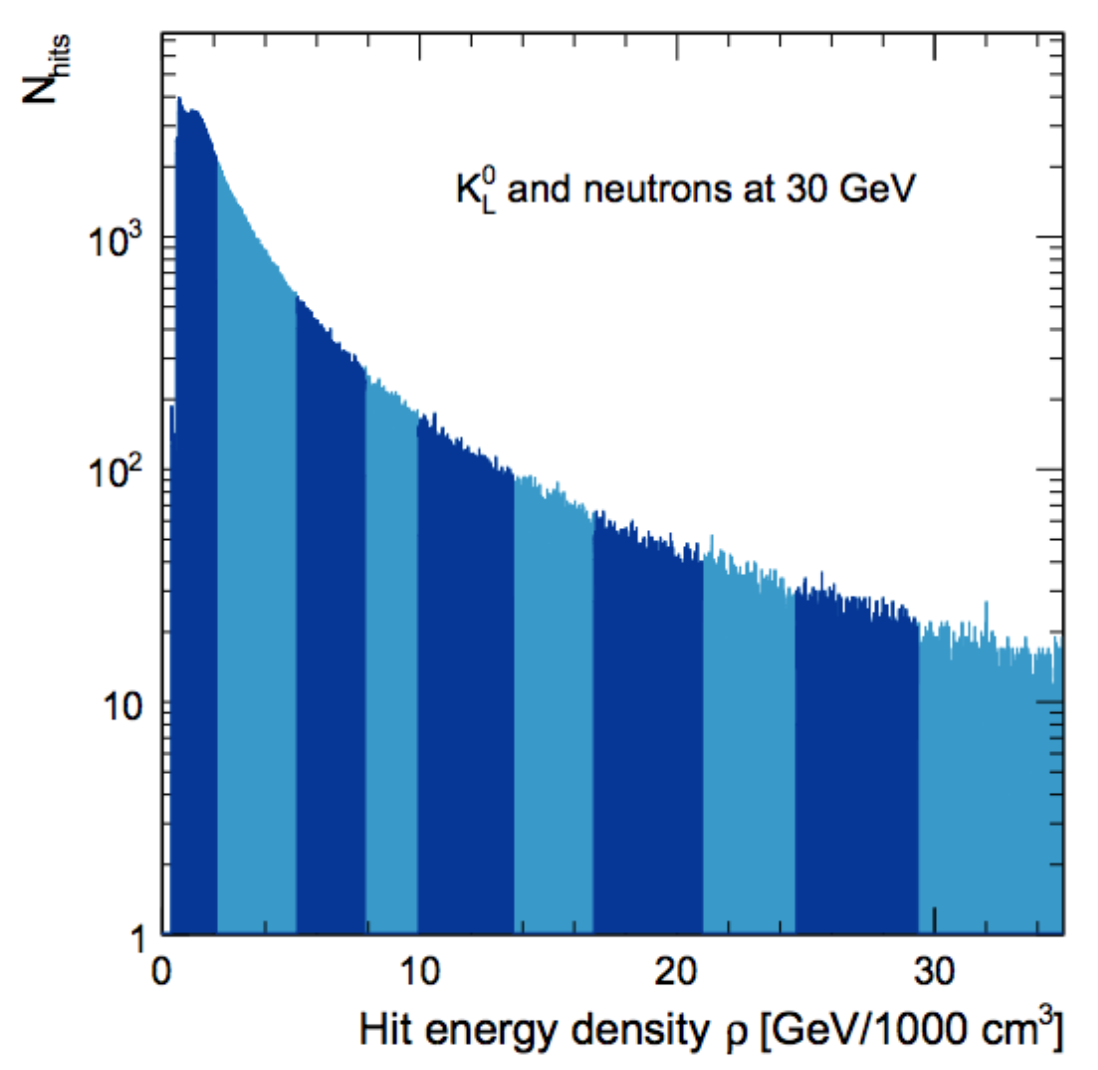}
  \caption{Energy density distribution for 30~GeV neutral hadrons. Differently colored areas are divided bins for software compensation \cite{SC}.}
  \label{figure:SCdens}
 \end{minipage}
 \hspace{0.04\hsize}
 \begin{minipage}{0.48\hsize}
  \centering
  \includegraphics[height=60mm]{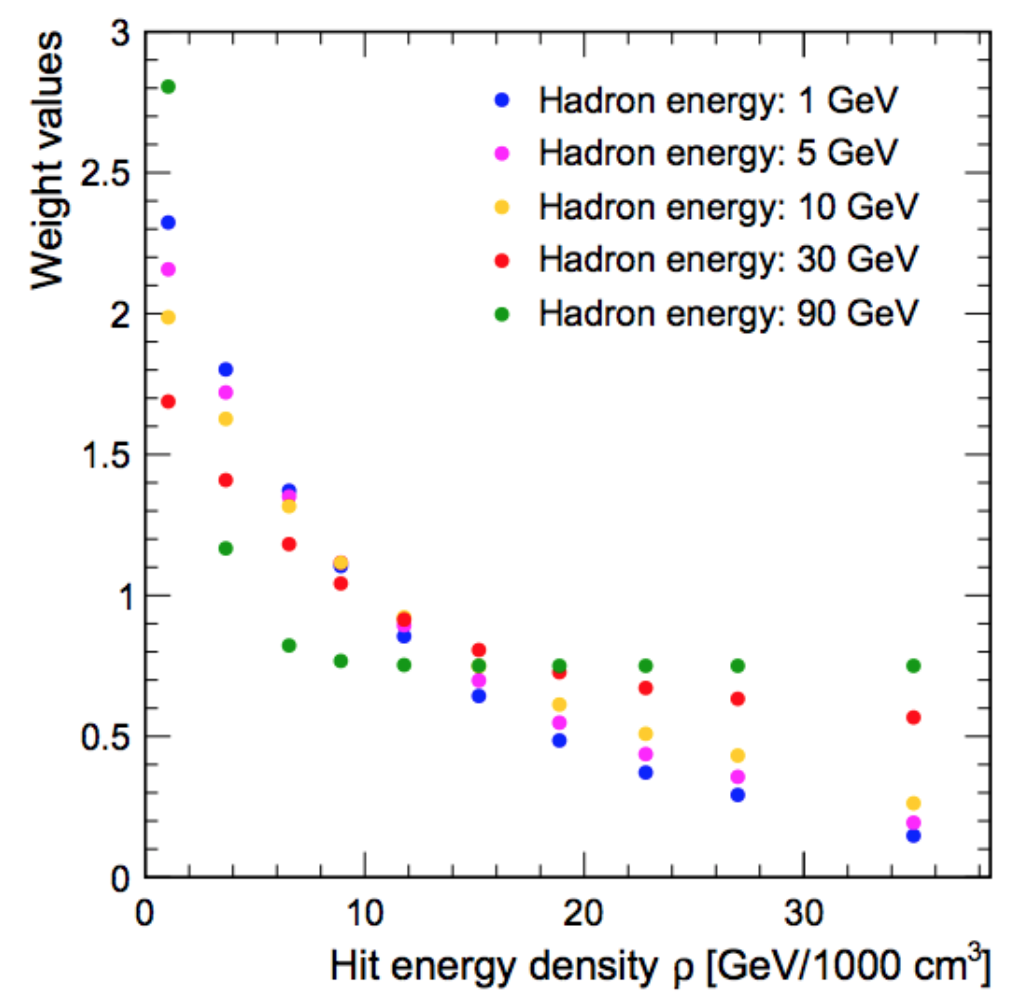}
  \caption{Standard weight values for different particle energies \cite{SC}.}
  \label{figure:SCweight}
 \end{minipage}
\end{figure}

Fig. \ref{figure:SCdens} shows a typical energy density distribution.
To define Eq. (\ref{eqn:weight}), the energy density is divided into 10~bins as illustrated in Fig. \ref{figure:SCdens}.
The sum of the hit energies in each bin is weighted by $w(\rho)$, with $\rho$ being the center of the bin.
The weight values for a standard ILD model is shown in Fig. \ref{figure:SCweight}.

\begin{figure}[h!]
 \begin{minipage}{0.48\hsize}
  \centering
  \includegraphics[height=60mm]{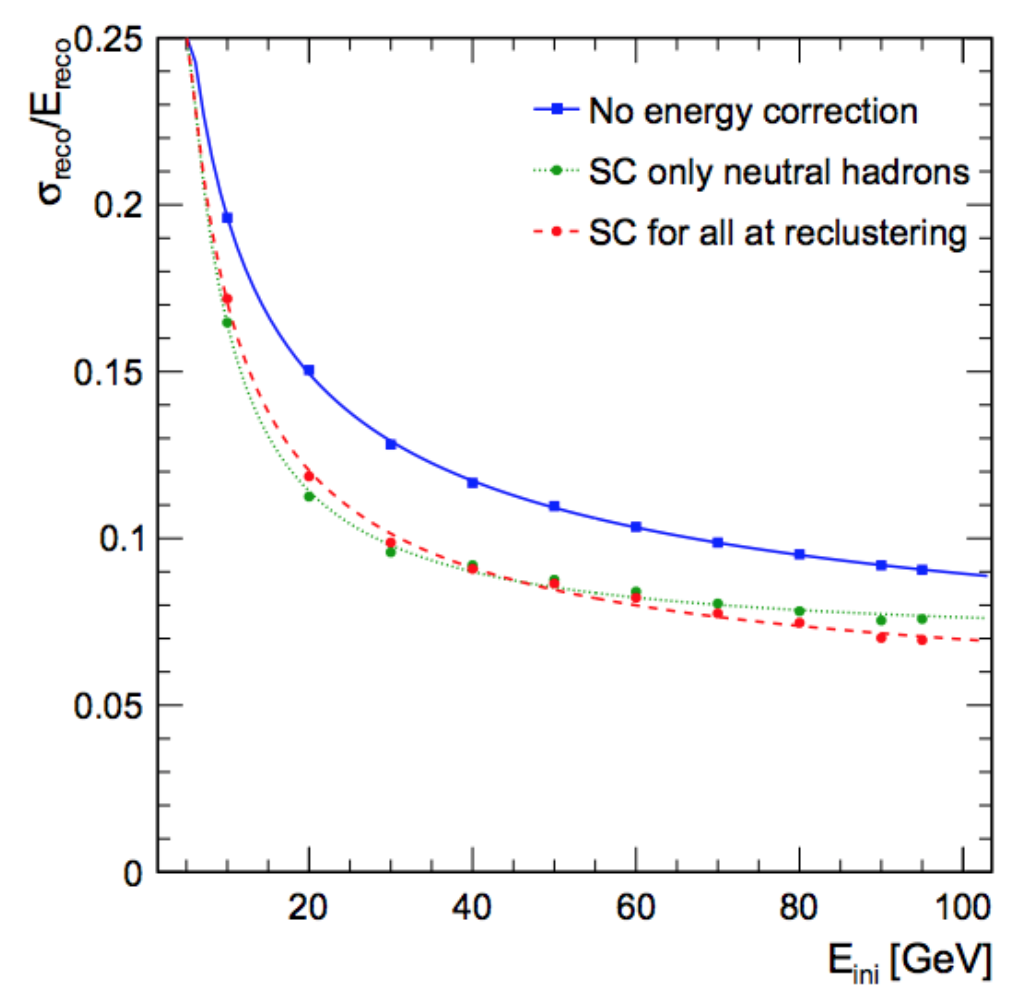}
  \caption{Energy resolution of single neutral hadron without energy correction (blue), applying SC only for neutral particles (green) and applying SC at re-clustering step for all clusters (red) \cite{SC}.}
  \label{figure:SCsingle}
 \end{minipage}
 \hspace{0.04\hsize}
 \begin{minipage}{0.48\hsize}
  \centering
  \includegraphics[height=60mm]{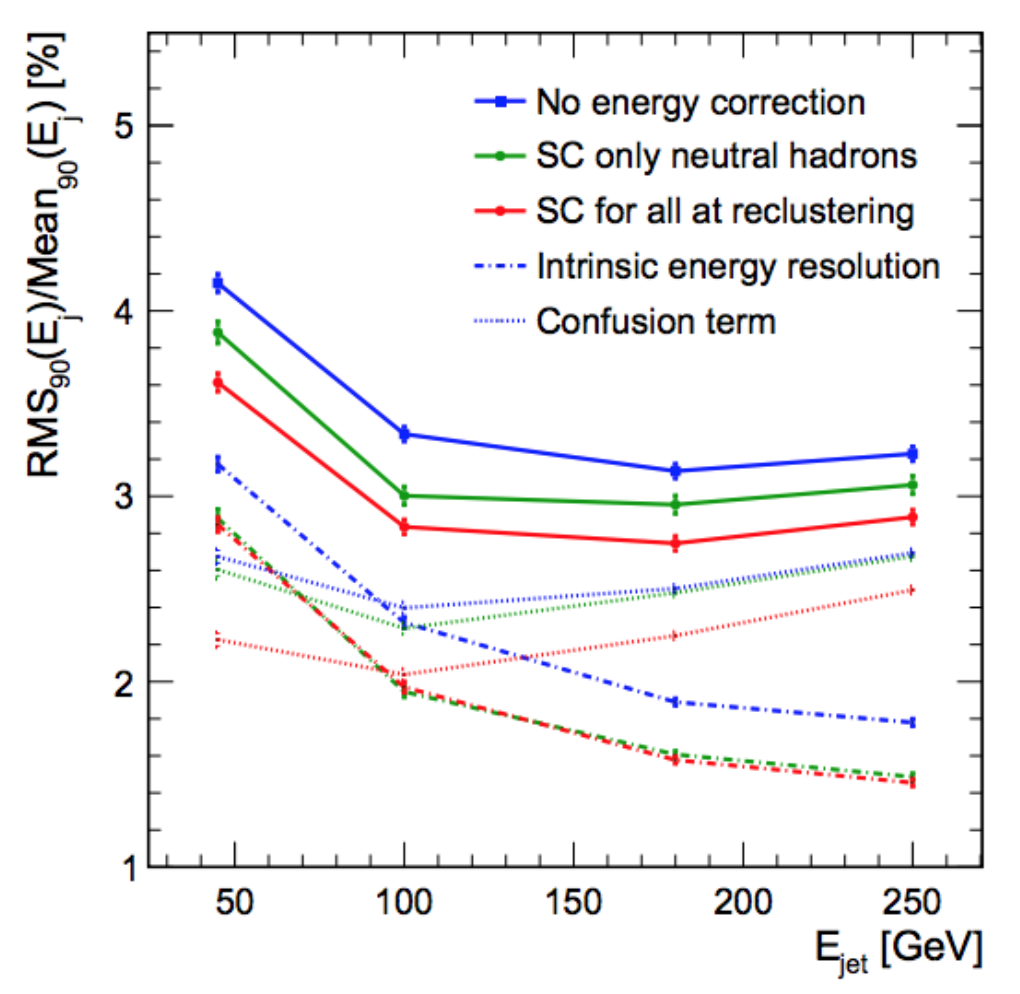}
  \caption{Jet energy resolution without energy correction (blue), applying SC only for neutral particles (green) and applying SC at re-clustering step for all clusters (red) \cite{SC}.}
  \label{figure:SCjet}
 \end{minipage}
\end{figure}

Introducing software compensation improves the energy resolution by 20\% as shown in Fig. \ref{figure:SCsingle} and Fig. \ref{figure:SCjet}.

Since the energy density distribution is changed by optical crosstalk, software compensation can be largely affected.
We performed the full re-optimization of the software compensation parameters using 10--100~GeV neutrons to mitigate the crosstalk effect.

\section{Results}
\begin{figure}[h!]
 \begin{center}
  \includegraphics[width=70mm]{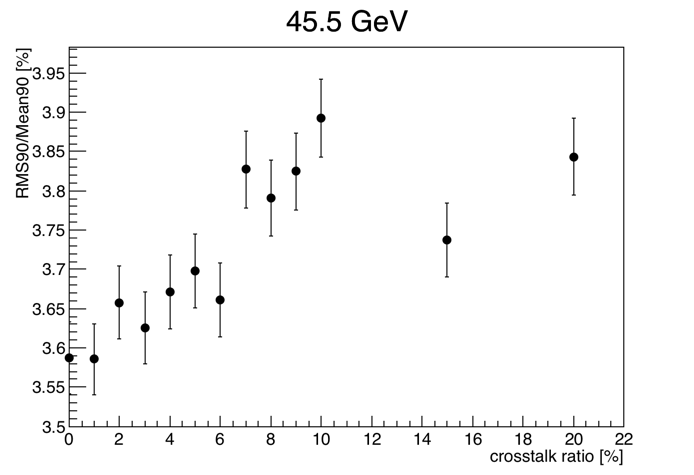}
  \includegraphics[width=70mm]{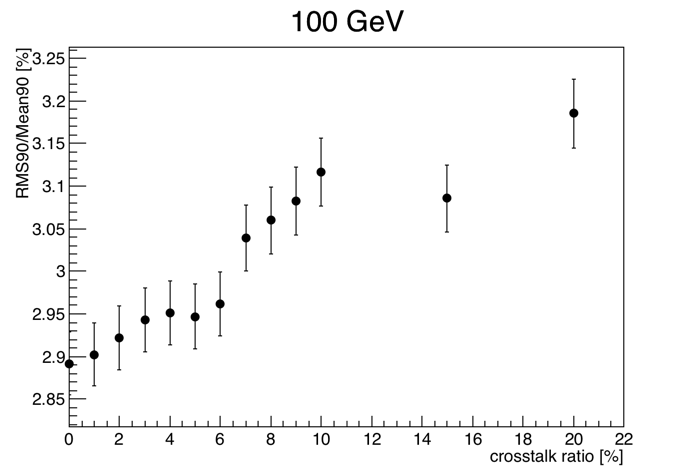}
  \includegraphics[width=70mm]{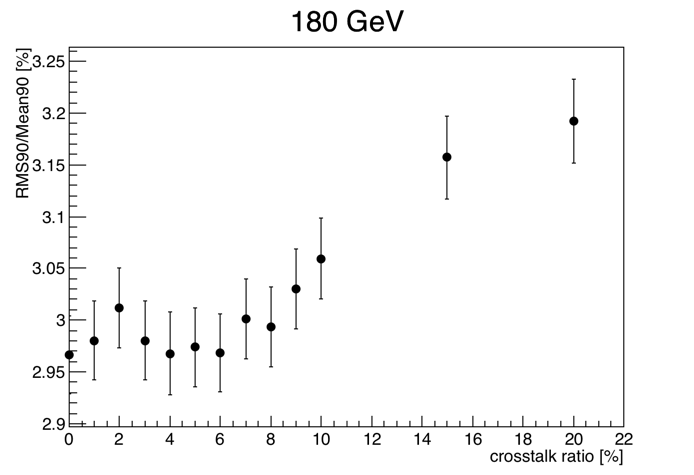}
  \includegraphics[width=70mm]{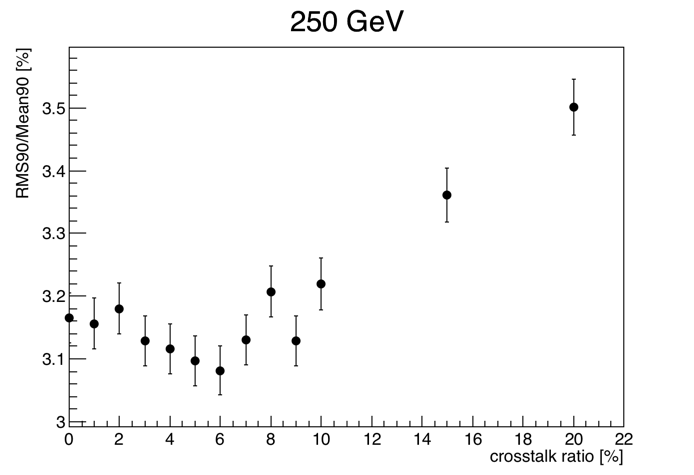}
 \end{center}
 \caption{Jet energy resolution as a function of the crosstalk fraction for different jet energies. The errors come from fit errors.}
 \label{figure:xtalkresult}
\end{figure}

\begin{figure}[h!]
 \begin{center}
  \includegraphics[width=110mm]{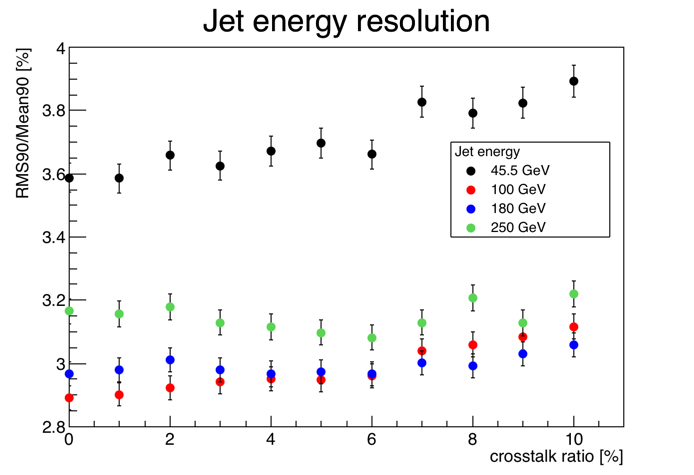}
 \end{center}
 \caption{Combined plot for Fig. \ref{figure:xtalkresult}.}
 \label{figure:xtalkresultall}
\end{figure}

After full calibration and re-optimization of the parameters for the Pandora PFA and the software compensation for each optical crosstalk fraction, the simulated jet events are reconstructed.
The mean value of the reconstructed jet energy didn't change by the crosstalk fraction.
Fig. \ref{figure:xtalkresult} and Fig. \ref{figure:xtalkresultall} shows the jet energy resolution as a function of the optical crosstalk fraction.

It can be seen that the low energy jets start to be affected even by a small optical crosstalk while the high energy jets as 250~GeV are not much affected by up to 4--6\% crosstalk.
A crosstalk of 5\% leads to a worsening of the energy resolution by 3\% for the 45.5~GeV jet.
A jet with low energy spreads in a wide range and leaked energy tend to be left undetected below the hit threshold.
This reduces the detection efficiency, hence worsens the energy resolution.
On the other hand, a jet with high energy is confined in a smaller space, most of the leaked energies are combined with neighboring hits and can be detected.

\section{Conclusions}
In some of the newly proposed active layer designs for the AHCAL, there might be a significant increase of the optical crosstalk between scintillator tiles compared to the baseline design.
However, the quantitative evaluation of the effect of the optical crosstalk had not been carried out.
We investigated the effect of the optical crosstalk on the jet energy resolution by a simulation study.

With full calibration and re-optimization of the reconstruction parameters, the mean reconstructed energy kept the same for every crosstalk fraction.
It was found that the effect of the crosstalk depends on the jet energy.
Low energy jets are quite sensitive even to small crosstalk while high energy jets are not much affected by up to 4--6\% crosstalk.
When the ILC is operated at low energy such as 250~GeV, the lower energy jets are quite relevant.
So the optical crosstalk should be minimized.

This simulation assumed a simple model where the optical crosstalk occurs only to the neighboring four tiles uniformly.
We plan to study with more realistic model including secondary crosstalk, the dependence of the crosstalk on the hit position in the tiles and so on after detailed measurements of the crosstalk with prototype active layers.

\section*{Acknowledgments}
We would like to thank all the CALICE and ILD colleagues.
Especially, we thank Daniel Jeans at KEK for his kind introduction to the iLCSoft,
Steven Green at Cambridge and Shaojun Lu at DESY for their detailed advises of the PFA calibration and analysis.

\bibliographystyle{unsrt}
\bibliography{LCWS2017}

\end{document}